\documentclass[fleqn,10pt]{wlscirep}
\usepackage[utf8]{inputenc}
\usepackage[T1]{fontenc}
\usepackage{float}
\usepackage{color, colortbl}
\usepackage[]{algorithm2e}

\title{Integrative Data Analytic Framework to Enhance Cancer Precision Medicine}

\author[1]{Thomas Gaudelet}
\author[1,2]{No\"{e}l Malod-Dognin}
\author[1,2,3,*]{Nata\v{s}a Pr\v{z}ulj}
\affil[1]{Department of Computer Science, University College London, London, WC1E 6BT}
\affil[2]{Barcelona Supercomputing Center (BSC), Barcelona, 08034 Spain}
\affil[3]{ICREA, Pg. Lluís Companys 23, 08010 Barcelona, Spain}

\affil[*]{natasha@bsc.es}

\begin{abstract}
With the advancement of high-throughput biotechnologies, we increasingly accumulate biomedical data about diseases, especially cancer. There is a need for computational models and methods to sift through, integrate, and extract new knowledge from the diverse available data to improve the mechanistic understanding of diseases and patient care. To uncover molecular mechanisms and drug indications for specific cancer types, we develop an integrative framework able to harness a wide range of diverse molecular and pan-cancer data. We show that our approach outperforms competing methods and can identify new associations. Furthermore, through the joint integration of data sources, our framework can also uncover links between cancer types and molecular entities for which no prior knowledge is available. Our new framework is flexible and can be easily reformulated to study any biomedical problems.

\end{abstract}
\begin{document}

\flushbottom
\maketitle

\thispagestyle{empty}

\section*{Introduction}

Over $18$ million new cases of cancer and 9 million deaths were recorded worldwide in 2018 \cite{international2018latest}. This makes cancer one of the leading causes of death. Cancer is a multi-faceted, complex disease arising from an accumulation of somatic mutations within the genome of normal cells that eventually leads to loss of normal cellular functioning and appearance of tumours that can spread across the body. Technological advances have enabled measurements from patient's tumour biopsies, including gene expression levels, DNA methylations, and somatic mutations. The research into cancer causes and treatments has greatly benefited from this wealth of patient data \cite{shen2015clinical,roos2019genomics}. 

Cancer projects, including The Cancer Genome Atlas (TCGA) and the International Cancer Genome Consortium (ICGC), have made publicly available wide-ranging, multi-modal, multi-omics cancer data, such as patient whole slide images, genome alterations, transcriptome, and epigenome \cite{tcga,icgc}. Free access to these large-scale, diverse databases has dramatically facilitated studies of the biological mechanisms of specific cancer types \cite{tcga,nik2016landscape,the2020pan}. The available data have also enabled pan-cancer analyses that study cancer in general to identify common mechanisms and differences across cancer types \cite{andor2016pan,the2020pan}. Recently, the Pan-Cancer Analysis of Whole Genome (PCAWG) project\cite{the2020pan} has informed that our knowledge about cancer is far from complete, as $5\%$ of their cohort was without any known cancer driver mutations.
Importantly, these large databases have paved the way for the field of Precision Medicine, whose aim is to improve medical care for patients by tailoring treatment to their individual molecular profiles \cite{ashley2016towards}. This is especially relevant to a heterogeneous disease, such as cancer, which manifests uniquely in every patient.

Complex diseases, such as cancer, can be caused by combinations of genetic, molecular, environmental, and lifestyle factors. Such diseases cannot be fully captured by any single type of biological data. As such, collective mining of different data has been gaining momentum as a means to extract integrated system knowledge that goes beyond what any single data source can offer \cite{ge2003integrating}. This principle applied to the study of cancer has enabled the discovery of cancer-related genes, or group of genes \cite{leiserson2015pan,gligorijevic2016patient,malod2019towards} and the identification of cancer sub-types significantly correlated with patient prognoses \cite{hofree2013network,gligorijevic2016patient}.

Biological data often have a small number of samples relative to the number of available features. For instance, a typical dataset in TCGA contains a few hundred patients that are each characterised by tens of thousands of features (e.g., expression levels of around 20,000 genes). However, biological features are often redundant due to underlying molecular interactions among biological entities \cite{kitano2002computational}. This has been a motivation for the use of dimensionality reduction and embedding algorithms that are pervasive in bioinformatics \cite{nelson2019embed}. Additionally, due to the low sample to features ratio, dimensionality reduction techniques are often necessary as data pre-processing for machine learning models\cite{nelson2019embed}. 

Non-negative matrix factorisation (NMF) approaches are unsupervised algorithms that have extensively been used both as a means to integrate heterogeneous data and to reduce data dimensionality. They encompass all methods that decompose a matrix, representing relational links between two sets of entities, into the product of low-dimensional, latent, positive matrices, or factors, whose sizes control the degree of dimensionality reduction \cite{wang2012nonnegative}. Importantly, they can be used to derive an embedding in an unspecified latent space for each entity. Matrix factorization approaches have had numerous applications, including collaborative filtering\cite{mehta2017review} and biological data integration for cancer analysis \cite{hofree2013network,gligorijevic2016patient}. Reconstructing a matrix based on a factorisation has often been used to make predictions and infer new knowledge \cite{gligorijevic2016patient}. However, unsupervised approaches often underperform supervised machine learning methods in classification tasks. Hence, NMF approaches have been combined with downstream machine learning classifiers \cite{guyot2018assessment}.

We propose a pan-cancer framework to uncover cancer type specific molecular mechanisms and identify drugs that could be re-purposed (see Figure \ref{fig:framework}). Our framework relies on the simultaneous integration and dimensionality reduction of various data using a joint non-negative matrix factorisation model. Our framework includes more data than the previous studies, integrating patient-specific diagnosis, gene expression, and single nucleotide variants as well as generic network data on human: protein--protein interactions, protein complex associations, biological pathways, drug--target interactions, and drug chemical similarities. To integrates the wealth of data in one framework, we rely on three types of matrix factorisations: non-negative matrix factorisation (NMF), non-negative matrix tri-factorisation (NMTF), and symmetric non-negative matrix tri-factorisation (SNMTF). The details for each can be found in Methods. Data integration is achieved by jointly optimising for multiple factorisation objectives with shared factors (see Methods). We obtain a \textit{context-aware} embedding of each entity (cancer type, patient, gene, complex, pathway, and drug) that takes into account all the input data. Using boosted decision trees, we predict biologically relevant associations between cancer types and genes, drugs, pathways, and complexes based on the context-rich embeddings of our entities. One key insight is that the integration step, by construction, embeds the entities into three latent spaces that contain: 1) patients and cancer types, 2) genes, complexes, and pathways, and 3) drugs. This means that the classifiers trained to predict associations for one entity can naturally be transferred to predict associations of another entity in the same latent space. In this respect, our approach is similar to zero-shot learning \cite{xian2017zero}, which aims to accurately classify at test time samples that belong to classes unseen at training time. In our case, we aim to predict the association of cancer types to unseen entities at training time. 

\section*{Results}
\subsection*{Our framework for context-aware embeddings}

\begin{figure}[H]
    \centering
\includegraphics[width=0.8\linewidth]{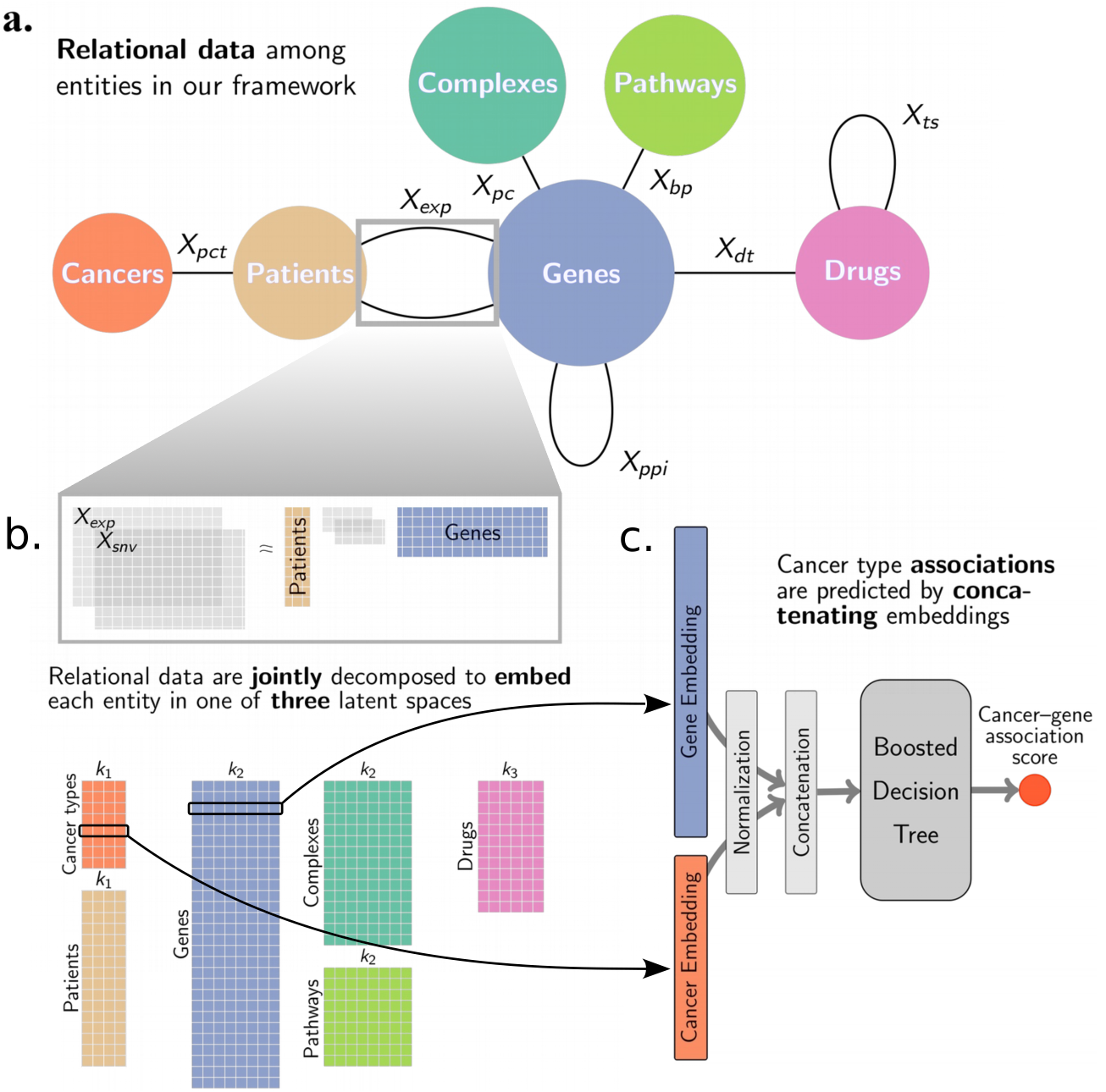}
    \caption{\textbf{a.} Input to our matrix factorization embedding model: relational data between entities. Each edge corresponds to a type of link and to a sub-objective of our joint factorisation model (see Methods Data Processing Table \ref{tab:data} for notation). The squares group entities that are embedded in the same joint latent space. \textbf{b.} Illustration of the NMTF factorisation sub-objectives corresponding to the edges in the grey box in panel a. Each group of entities is associated in the decomposition to a factor that is shared across all sub-objectives involving that group of entities. Through the joint decomposition of all relational data, we derive embeddings for each entity in three latent spaces with dimensions $k_1$, $k_2$, and $k_3$. \textbf{c.} We predict associations relevant to cancer types with boosted decision tree classifiers taking as input, for instance, the concatenation of the embeddings of a cancer type and of a gene.}
    \label{fig:framework}
\end{figure}

The core of the framework is gene information (see Figure \ref{fig:framework}.a.). We integrate three types of data about genes. We obtain RNA sequencing (RNA-seq) data and single nucleotide variants (SNV) data for 7,998 patients from ICGC (see Methods for details) across $21$ cancers. Henceforth, we refer to each cancer by its abbreviation given in Table \ref{tab:cancers}. We obtain data on gene interactions including protein-protein interaction (PPI) network from Biogrid, protein complexes (PC) from Reactome and Corum, and biological pathways (BP) from Reactome (see Methods for details). These data capture physical and functional relationships between genes and are used to anchor our framework within the context of molecular interactions. The last type of gene data corresponds to drug--target interactions from DrugBank (see Methods for details), connecting drugs to proteins that they target. We further add drug chemical similarity information to push similar drugs closer in the latent space. We also add patient diagnosis information through which we embed cancer types and patients in a joint latent space to both push patients closer if they have the same cancer and push molecularly similar cancer types closer. This could help tailor treatments to patients by placing them within a cancer ``space'' since cancer is a heterogeneous disease and a given cancer type might manifest differently in different people. This may aid characterising cancer of each patient as accurately as possible to personalise treatment options. 

Because of the heterogeneity of our input data, our integration framework is based on joint optimisation of different variants of non-negative matrix factorisation: classical Non-negative Matrix Factorisation (NMF), Non-negative Matrix Tri-Factorisation (NMTF), and Symmetric Non-negative Matrix Tri-Factorization (SNMTF). Each variant, described in Methods, is best fitted for the decomposition of a different type of relational data. In particular, we use SNMTF to factorise the PPI network and the drug similarities matrix, NMTF to factorise patient molecular data and drug--target data, and NMF for the remaining data. Each edge in Figure \ref{fig:framework}.a corresponds to a sub-objective of our embedding framework, i.e. a specific NMF decomposition. Each group of entities is associated in the joint decomposition to a factor that is shared across all sub-objectives involving that group of entities. For instance, the patient factor is shared by all sub-objectives that involves patient-specific data (diagnoses, gene expressions, and somatic mutations). An entity's embedding is obtained from the factor of the associated group of entities.

Through our integrative framework, we derive embeddings for all entities (cancer types, patients, genes, pathways, complexes, and drugs) that best fit the full context of the framework, i.e. the input relational data. Each entity's embedding, in one of the three latent spaces learnt by our framework, encapsulates the information from the input data that is relevant to that entity; thus we say that this representation is  \textit{context-aware}. Our framework has three hyperparameters, denoted by $k_{1}$, $k_2$ and $k_3$, which correspond to the dimensions of the latent spaces. To find suitable values for these hyperparameters, we perform a grid search with $k_{1} \in \{2,5,10,15,21\}$, $k_2\in\{70,80,90,100,110\}$, and $k_3\in\{40,50,60,70,80\}$. As the selection criterion, we measure if each patient tends to be embedded in the latent space closer to their diagnosis than to other cancer types. We quantify this with the macro-F1 score of the classifier that associates to each patient the closest cancer type in the latent space 

\noindent in terms of cosine distance. We found that the following hyperparameters values maximize this metric: $k_1=21$, $k_2=70$, and $k_3=40$. Supplementary Figure 1 shows the sensitivity of different metrics to the choice of the hyperparameters, which we discuss in the rest of the article.

 \begin{figure}[ht]
    \centering
\includegraphics[width=0.9\linewidth]{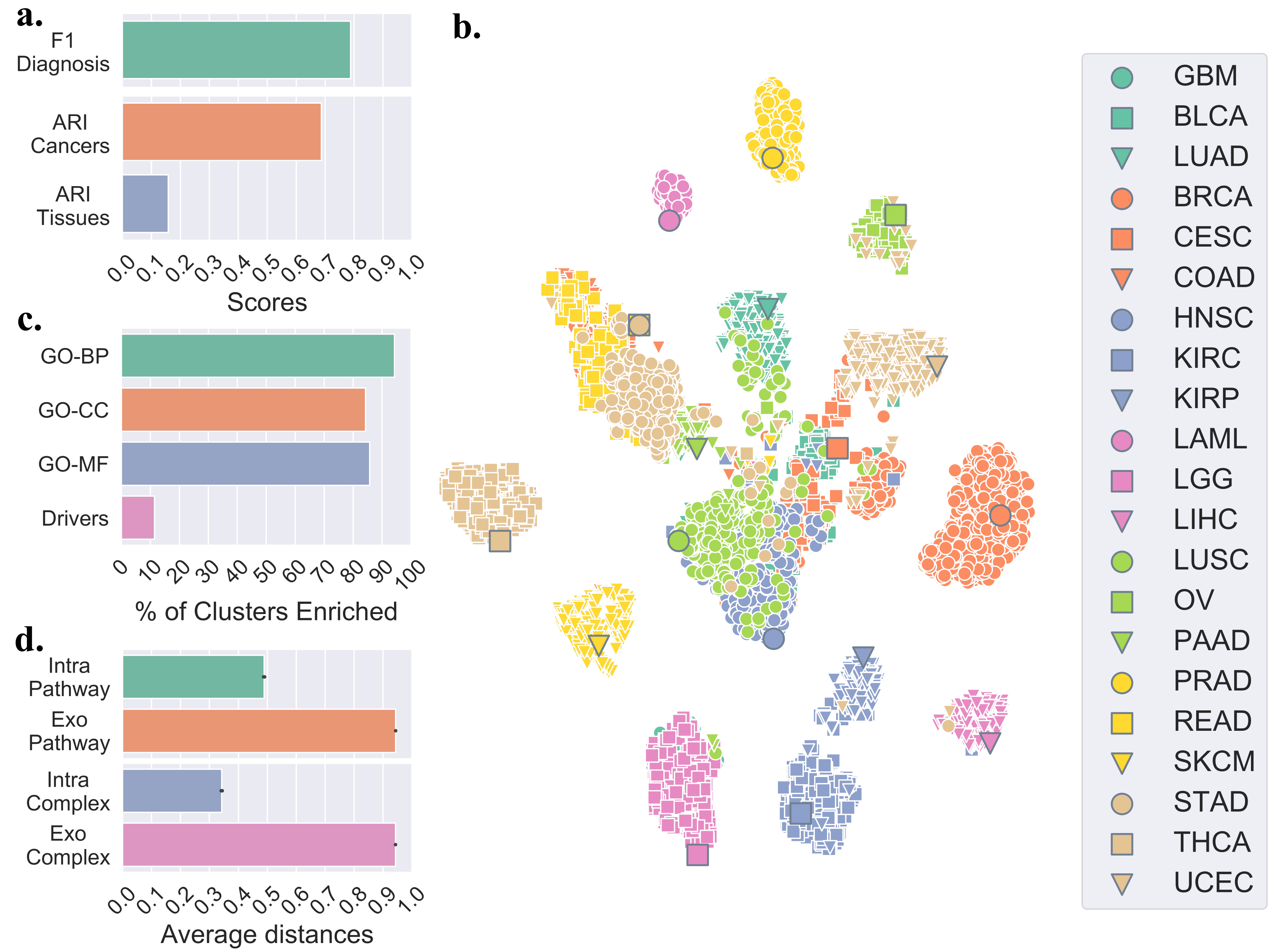}
    \caption{\textbf{a.} Macro-F1 score quantifying the relation between a patient and its cancer type (green) and adjusted random index (ARI) measuring the link between patient clustering for each model and cancer type labelling (orange) and tissue sampled labelling (blue). \textbf{b.}  t-SNE plot representing the embedding of patients and cancer types in the latent space. The larger circled markers correspond to the embeddings of cancer types, and the smaller ones represent the embeddings of patients. Colours and shapes indicate cancer types (see Methods Table \ref{tab:cancers} for abbreviations meanings). \textbf{c.} Percentages of gene clusters enriched in  GO-BP, GO-CC, GO-MF, and driver annotations. \textbf{d.} Average cosine distance between genes and associated pathways and complexes (intra-pathway and intra-complex) and non-associated pathways and complexes (exo-pathway and exo-complex).}
    \label{fig:embedding}
\end{figure}

\subsubsection*{Patient and cancer embeddings are medically relevant}
To evaluate the biomedical relevance of our joint patient and cancer embeddings, we observe that the macro-F1 score is close to $0.8$ for our optimal set of hyperparameters (see Figure \ref{fig:embedding}.a.) indicating that the majority of patients are embedded closer to their diagnoses than to other cancer types. In addition, we evaluate if patients group in the latent space with respect to either cancer type, or a sampled tissue. To this end, we use hierarchical clustering with cosine distance to group patients in $k$ groups (where $k$ is either the number of cancers, or the number of tissues) and compute the Adjusted Rand Index (ARI) to measure the link between the clustering and the ground truth labelling (either cancer types, or sampled tissues; see Figure \ref{fig:embedding}.b.). We observe that patients do not cluster well with respect to sampled tissues, having ARI below $0.2$. However, we observe ARI $0.7$ with respect to cancer type, indicating that our clusterings resemble diagnostic labelling with some discrepancies. These results are expected, as the inclusion of patient diagnosis data in the framework implies a constraint that aims to embed each patient close to their diagnosis and subsequently, to other patients having the same disease. Note however, that some patients do not fit well with the rest of their cohorts. This is an important observation, as it suggests that those patients might need different care options from the majority of their cohort and further motivates personalising treatments to individual patients.

As an illustration, we visualise our latent space embeddings using T-distributed Stochastic Neighbor Embedding (t-SNE). t-SNE is a machine learning algorithm for nonlinear dimensionality reduction, well-suited for visualisation in a two-dimensional space of high-dimensional data  \cite{maaten2008visualizing}. We observe as expected that patients tend to cluster according to cancer type, with the cancer type itself being also embedded nearby (see Figure \ref{fig:embedding}.c.). Additionally, we observe that some cancers are grouped in a meaningful way. For instance, both brain cancers, GBM and LGG, form one group. The cluster in the centre contains mostly squamous cell carcinomas, HNSC, CESC, and LUSC. Both cancers that affect kidneys, KIRP and KIRC, are also grouped. And READ, COAD, and STAD - which are cancers affecting rectum, colon, and stomach, respectively - form another cluster. We also observe that some patients having a certain type of cancer do not group with the majority of the cohort. 

We further investigate if our framework learns a meaningful latent space that translates into actionable representations for new unseen patients. We perform a $10$-fold cross-validation in which $90\%$ of patients of each cancer type are used to derive the embeddings of all entities in the framework. We then project the remaining $10\%$ of patients in the derived cancer/patient latent space (see Supplementary Methods). First, we test if these patients are placed close in space to their diagnosis (quantified as above, see Supplementary Table 1). This gives a macro-F1 score of $0.77 \pm 0.03$, which is close to the score obtained with all patients included in the framework ($\sim 0.8$). This shows that new patients are placed in the latent space according to their diagnoses with accuracy similar to that for patients included in the framework. We also test if the unseen patients tend to be embedded closer to patients having the same diagnoses. For this, we use a k-nearest neighbours classifier with $k=10$ (see Supplementary Methods) and measure its macro-F1 score (see Supplementary Table 1). We observe a score of $0.88 \pm 0.02$, which shows that the large majority of new patients are embedded in the latent space closer to patients diagnosed with the same type of cancer. Both results show that our latent space is robust in the sense that we can derive an embedding for new patients that is consistent with that of known patients and cancer types. We also observe that the k-nearest neighbour algorithm gives a more robust diagnosis classifier than finding the closest cancer in the latent space. This means that the local neighbourhood of a patient in the latent space is a better diagnosis indicator than a global predictor derived from cancer types' embedding. This suggests the presence of patient subgroups within a cancer type that display substantially different molecular behaviour. 

Overall, our analysis shows that the patients/cancers latent space is consistent with known biology. Furthermore, our framework has the advantage of relaxing the hard clustering derived from patients diagnoses through patient's molecular similarity, highlighting that a patient's molecular profile can be more similar to the profiles of patients with different cancers than to the profiles of patients with the same diagnosis. This observation motivates further the need for pan-cancer perspectives in precision medicine.

\subsubsection*{Our gene latent space is biologically relevant}
To evaluate the biological relevance of our genes embeddings, we cluster them in $k_2$ group and measure the enrichments of the clusters in terms of Gene Ontology (GO) annotations and in terms of cancer driver genes  (see Methods for details). We consider all three subtypes of GO annotations: Biological Processes (GO-BP), Cellular Component (GO-CC), and Molecular Function (GO-MF) separately. The significance of the enrichments is computed with a hypergeometric test with Benjamini-Hochberg correction for multiple hypothesis testing and a significance threshold of $0.05$. We observe that, regardless of the GO subtype, above $80\%$ of clusters are significantly enriched in at least one annotation (see Figure \ref{fig:embedding}.c.). These results show that genes with similar function are embedded closer in the latent space and thus that our genes' embeddings capture known biology. Interestingly, we also observe that around $10\%$ of the clusters are enriched in cancer driver genes, indicating that cancer drivers are embedded closely, i.e. clustered, in the latent space. This highlights the link between the gene latent space and the cancer context that we made a part of our framework. Furthermore, it underlines the relevance of our embeddings for the identification of putative cancer-related genes, discussed in the following section.

Finally, as pathways and complexes are embedded in the same latent space, we investigate their positioning with respect to genes. In particular, we evaluate if a gene is embedded closer to its associated higher-order entities, i.e. pathways and complexes, than to those to which it has not been associated yet. To this end, we compute the cosine distances in the latent space between a gene and its associated pathways and complexes, termed ``intra--pathway'' and ``intra-complex'' distances, as well as the distances between the gene and all non-associated pathways and complexes, termed ``exo--pathway'' and ``exo-complex'' distances.  We observe that genes are embedded closer, on average, to their associated higher-order entities than they are to those that they are not associated to (see Figure \ref{fig:embedding}.d.), with average distance below $0.5$ between a gene and associated entities and above $0.9$ between a gene and non-associated entities. These results are significant according to a Mann--Whitney U statistical test (p-value $\sim 0$ in both cases) and underline the relevance of the joint embedding of genes with related higher-order molecular structures in the same latent space. This also suggests that our framework could be used for identifying new genes that are involved in, or interact with pathways and protein complexes, which we leave for future work.

\subsection*{Predicting cancer type associations}
To extract new knowledge for each cancer type, we use our context-aware embeddings to suggest cancer--drug and cancer--gene associations. In particular, we train classifiers to predict known associations based on our entities' embeddings and use the trained classifiers to predict new associations (see Methods). We first normalise all embeddings to have a unit norm. The normalisation step is crucial for the transfer of a classifier between types of entities that we discuss in the following section. Then, for each cancer--drug pair (or cancer--gene pair), we define the pair's representation by the concatenation of the embeddings of its components, i.e. the concatenation of the cancer's embedding vector with the drug's embedding vector defines the feature vector of the pair. Finally, we use boosted decision trees as simple classifiers that take as input a pair's representation and output the association's scores of its component (see Figure \ref{fig:framework}.c.). We choose boosted decision trees due to their simplicity and high performances in a number of competitions \cite{chen2016xgboost}.

In the first validation step, we systematically evaluate the performance of our approach with a $10$-fold cross-validation using both the Area Under the Receiver Operating Curve (AUROC) and the Area Under the Precision recall Curve (AUPRC) and compare our results to state-of-the-art predictive methods. In the second step, we investigate the top $10$ drugs and genes associated to cancer types by our methodology. Each pair is scored based on the average of the standardised scores given by $10$ classifiers trained for the cross-validation (see Methods for details). In this step, we only consider drugs and genes that were thus far not associated to any cancer type in the ground-truth data (introduced in each subsection) to avoid trivial cases 
\begin{figure}[ht]
\centering
\includegraphics[width=\linewidth]{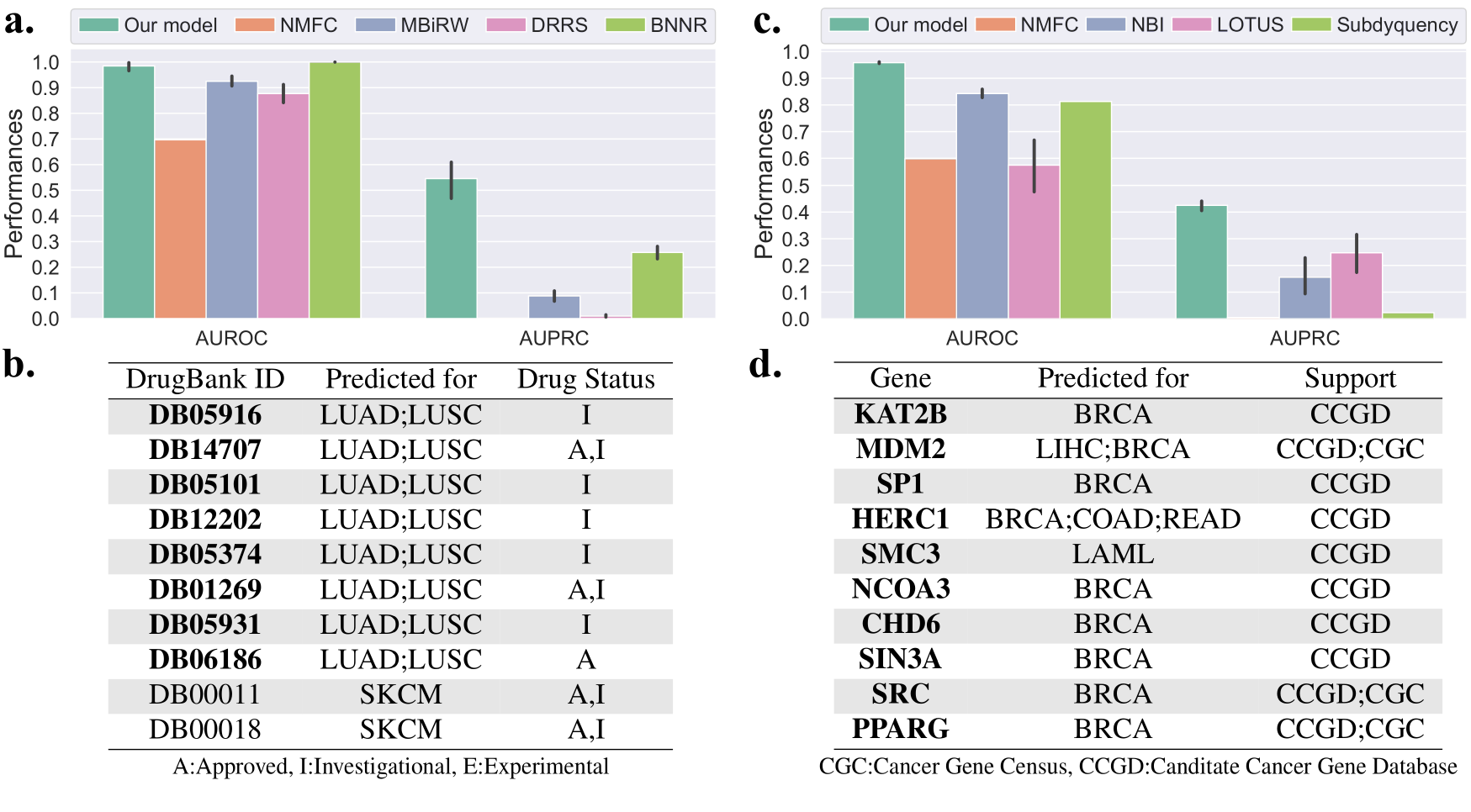}
    \caption{Performances of our cancer--drug association predictor (left column) and cancer--gene associations predictor (right column). The bar charts give the performances of our classifiers measured with 10-fold cross-validation in terms of Area Under the Receiver Operating Characteristic (AUROC) and Area Under the Precision Recall Curve (AUPRC). The tables give the top ten associations between cancers and drugs (panel b.) and genes (panel d.) that are not associated with any cancers in our data. Drugs or genes highlighted in bold font have been associated to cancer. The support column in the bottom right table indicates which database lists a link between the gene and cancer.}
    \label{fig:predictions}
\end{figure}{}
\noindent of information transfer from one cancer to another, which typically happens when one drug, or one gene is associated to a majority of cancers. We perform a manual literature curation to validate the top results.

\subsubsection*{Our model predicts relevant treatment}
To predict cancer--drug associations, we train classifiers to identify known associations that we collect from DrugCentral\cite{ursu2016drugcentral} (last updated October 2018). We define our positive set with DrugCentral treatment options and consider all non-reported associations for our negative set. Our classifier takes as input the concatenation of the normalised embeddings of a drug and a cancer type and outputs their association score. We compare our results to four baseline methods: Non-negative Matrix Factorization Reconstruction (NMFR),  Measure-based Bi-directional Random Walks (MBiRW) \cite{luo2016drug}, Drug Repositioning Recommendation System (DRRS)\cite{luo2018computational}, and Bounded Nuclear Norm Regularization (BNNR)\cite{yang2019drug} (see Methods for implementation details).  

We observe that our approach significantly outperforms the competing methods (see Figure \ref{fig:predictions}.a.). BNNR achieves slightly better AUROC scores ($\sim 0.99$ compared to our $\sim 0.97$), but it scores significantly lower than our framework in terms of AUPRC ($\sim 0.25$ compared to our $\sim 0.5$). These results show the relevance of our method when compared to the state-of-the-art drug re-purposing approaches. We analyse further the results of our approach through literature curation for the top-scoring drugs that are not associated with any cancer types in DrugCentral.

Among the top $10$ drugs that are the most associated to cancer types by our classifiers (see Figure \ref{fig:predictions}.b.), a majority is recorded in DrugBank as investigational, or approved for the treatment of some cancers. The approved predicted drugs either are not present in DrugCentral, as their approval postdates the DrugCentral release, or target a cancer type not considered in this study. We discuss below supporting information for our top $3$ predicted drugs. We provide validations of all of our predictions in the Supplement.
 
DB05916 (CT-011) targets gene PDCD1, which has immunomodulating and antitumor activities. CT-011 is currently being investigated for the treatment of tumours and unspecified cancers \cite{wishart2017drugbank}. DB14707 (Cemiplimab) is an FDA approved drug for the treatment of advanced cutaneous squamous cell carcinoma \cite{wishart2017drugbank}. Our classifier suggests that it could be used to treat lung cancer and notably lung squamous cell carcinoma (LUSC). DB05101 (Matuzumab) is an investigational drug that targets EGFR gene, which is often associated with cancers, including lung cancers\cite{sos2009pten}. 

The manual literature curation highlights that our predicted drugs are often investigated, or approved for the treatment of forms of cancer and that their targets, or mechanisms of actions, can be linked to the specific cancer types we predict. Overall, the analysis strongly supports our methodology.

\subsubsection*{Our framework identifies genes relevant to cancer types}
Based on known cancer genes from IntOGen \cite{gonzalez2013intogen}, we train classifiers to identify associations between genes and cancer types. Our positive set is constituted of cancer--driver associations. All non-reported associations are considered as part of our negative set. As above, our classifier takes as input the concatenation of the normalised embeddings of a gene and a cancer type and outputs their association score. We compare the performance of our method with the following state-of-the-art methods: Non-negative Matrix Factorisation Reconstruction (NMFR), Network Based Integration (NBI)\cite{ruffalo2015network}, LOTUS \cite{collier2019lotus}, and Subdyquency\cite{song2019random}. The methods were developed to predict cancer-related genes in slightly different contexts and are adapted to our problem here (see Methods for details).

We observe that our approach outperforms the competing methods (see Figure \ref{fig:predictions}.c.) in terms of AUROC, which is over $0.9$ for our method compared to below $0.8$ for the other approaches, and in terms of AUPRC which are around $0.4$ for our approach compared to below $0.25$ for the other methods. We further look at the top $10$ genes that are identified by our method (see Figure \ref{fig:predictions}.d.). We use the Cancer Gene Census (CGS)\cite{sondka2018cosmic} and Candidate Cancer Gene Database (CCGD) \cite{Abbott2015a} to find known associations, as well as literature curation (both databases were accessed in August 2019).

We observe that our top $10$ scoring genes are listed in either the Candidate Cancer Gene Database (CCGD), or the Cancer Gene Census (CGC) as linked to at least one form of cancer. Furthermore, the pairs MDM2--LIHC, HERC1--COAD, HERC1--READ, SMC3--LAML, NCOA3--BRCA, and CHD6--BRCA are associated in CCGD. Additionally, KAT2B (PCAF) activity has been linked to cancers, and in particular, to breast cancers, in the literature \cite{zhao2014notch,zhang2017microrna,bondy2019nonhistone}. MDM2 has also been associated with breast cancer \cite{lukas2001alternative}. SP1 expression has been linked to breast cancer in multiple prior studies \cite{duan1998estrogen,wang2007expression,wang2017hbxip}. For each of these three genes, we stratify our BRCA cohort into two groups: patients having higher than average expression of the gene and patients having lower than average expression of the gene. We compute a logrank statistical test (with $0.05$ cut-off) and observe that for each of the three genes, the patient groups have significantly different survival rates with p-values $0.002$ for KAT2B, $0.026$ for MDM2, and $0.039$ for SP1 (see Supplementary Figure 2.a-c. for Kaplan-Meir plots). For each of the three genes, higher expressions are associated with lower survival rates. We provide validations for the remaining predictions in the Supplement.

The literature curation highlights that each gene identified through our approach is relevant to the associated cancer type, with support through existing research and databases, as well as statistical evidence for a connection between gene expression level and patient prognosis. Thus, this supports our methodology.

\subsection*{Re-purposing classifiers}

Links between types of entities are not always known or available (e.g., associations between cancer type and protein complexes or associations between patients and drugs), which prevents us from using the same methodology to derive new knowledge. However, our framework allows for extrapolating those links from known associations between other types of entities. Our approach relies on the previously observed fact that, by design, some entities are embedded in the same latent spaces (e.g., genes, pathways, and complexes or cancer types and diseases). We have further shown in the first section, that the relative location in the latent space of related entities was biologically consistent, i.e. related entities are closer to each other than non-related entities. Based on these observations, we postulate that a classifier trained from the embeddings of a given type of entities can be re-purposed to predict from the embeddings of another type of entities. For instance, classifiers that learnt to associate genes to cancer types can be used to predict which biological pathways or protein complexes could be associated with which cancer types. This could effectively provide mechanistic insights into the inner workings of cancers by identifying affected higher-order cellular structures and functions. Furthermore, by pushing this logic further, one can derive patient-specific predictions, particularly patient-specific drugs. This is, however, difficult to validate due to the lack of patient treatment information in ICGC database. Thus, we focus below on the analysis of top associations between higher-order cellular structures and cancer types.

\subsubsection*{Our re-purposed classifiers identify cancer-related protein complexes}
To obtain the association score between a cancer type and a protein complex, we simply feed the concatenation of the normalised embeddings of both entities to the $10$ classifiers trained to predict cancer--gene associations. The average of the standardised scores across all classifiers gives the final association score.

To the best of our knowledge, there are no comprehensive database reporting associations between cancers and protein complexes. Thus, we are unable to provide global validation scores for our predictions. We proceed by validating the top $3$ scoring protein complexes manually (see Table \ref{tab:complex}) below. We provide validations of all of our predictions in the Supplement.
\begin{table}[ht]
      \centering
      \begin{tabular}{c l }
      \hline
          Protein Complex  & Predicted for \\
          \hline
            \rowcolor{black!10}
         IL6:sIL6R:IL6RB:JAKs & all cancers\\
         p-7Y-RUNX1:PTPN11 & BRCA;BLCA;PAAD;LUSC;GBM   \\
            \rowcolor{black!10}
         R-HSA-1112759 & BRCA;LAML;BLCA \\
         Integrin alpha2bbeta3:SRC & BRCA;BLCA  \\
            \rowcolor{black!10}
         R-HSA-1112753 & BRCA;LAML;BLCA;LGG;GBM  \\
         R-HSA-1112563 & BRCA   \\
            \rowcolor{black!10}
         SAM68:p120GAP & BRCA  \\
         JAKs:OSMR & BRCA  \\
            \rowcolor{black!10}
         IL6ST:JAKs & BRCA \\
         R-HSA-9632399 & BRCA \\
           \hline
      \end{tabular}{}
    \caption{Top $10$ protein complexes associated to cancer types.}
     \label{tab:complex}
\end{table}

IL6:sIL6R:IL6RB:JAKs complex plays a role in interleukin 6 signalling, which is linked to cancer \cite{kumari2016role}. The complex is associated with JAK family of kinases, themselves tied to cancer \cite{verma2003jak}. p-7Y-RUNX1:PTPN11 complex is involved in the regulation of RUNX1 expression and activity. RUNX1 has been linked to various cancer, sometimes with opposite effects \cite{Abbott2015a}. However, regardless of its precise role in a given cancer, the regulation of RUNX1 appear to be of critical importance as over- or under-expression can have an important impact on the development of cancer\cite{janes2011runx1}.  Tyrosine phosphorylated IL6 receptor hexamer:Activated JAKs:Tyrosine/serine phosphorylated STAT1/3 complex (R-HSA-1112759) is involved with interleukin 6 signalling and more specifically serine phosphorylation of STAT family of transcription factors. We have seen previously that interleukin 6 signalling has been linked to cancer. Furthermore, STAT has been linked to various cancers, including breast cancer (BRCA) that our results associate to the protein complex \cite{clevenger2004roles}. 

The literature curation highlights that our predicted associations between protein complexes and cancer types are supported by the existing literature. Thus, the analysis demonstrates the validity of our classifier re-purposing approach, as well as the ability of our framework to extract cancer mechanisms at the level of protein complexes.

\subsubsection*{Our re-purposed classifiers identify cancer-related biological pathways}

Similarly, we can predict associations between cancer types and biological pathways. The association score between a cancer type and a biological pathway is obtained by simply feeding the concatenation of the normalised embeddings of both entities to the $10$ classifiers trained to predict cancer--gene associations. The average of the standardised scores across all classifiers gives the final association score.

CTD database \cite{ctd_database} gives associations between diseases and pathways based on shared associated genes and can be used for global validation. We achieve an AUROC score of $0.65\pm 0.01$ and an AUPRC score of $0.66 \pm 0.01$, which indicates predictions significantly better than random (p-value $\sim 0$) for our re-purposed classifiers. However, note that $52\%$ of all possible associations between our set of cancer types and our set of pathways are reported in the database. This indicates that the condition for association used by CTD might not be sufficiently stringent. This motivates the following manual literature curation to validate our top $10$ scoring biological pathways. We discuss the first $3$ predicted pathways below and provide validations of all remaining predictions in the Supplement.

\begin{table}[ht]
\centering
      \begin{tabular}{c p{5.75cm}}
      \hline
          Pathway & Predicted for \\
          \hline
            \rowcolor{black!10}
          R-HSA-112411 & all cancers\\
          R-HSA-2262752 & BRCA\\
            \rowcolor{black!10}
          R-HSA-5654688 & BRCA;GBM;BLCA;LGG;PAAD;LAML; LUSC;STAD;SKCM;LUAD;UCEC \\
          R-HSA-5654699 & BRCA \\
            \rowcolor{black!10}
          R-HSA-8953897 & BRCA\\
          R-HSA-110056 & BRCA \\
            \rowcolor{black!10}
          R-HSA-389357 & BRCA\\
          R-HSA-5654719 & BRCA  \\
            \rowcolor{black!10}
          R-HSA-9603381 & BRCA \\
          R-HSA-8866910 & BRCA\\
           \hline
      \end{tabular}{}
    \caption{Top $10$ biological pathways associated to cancer types.}
    \label{tab:pathway}
\end{table}{}

MAPK1 (ERK2) activation pathway (R-HSA-112411) and MAPK3 (ERK1) activation pathway (R-HSA-110056) have been linked to numerous cancers, such as breast cancer, as discussed in the previous section, and colorectal cancer \cite{fang2005mapk}. The ERK MAPK pathway is critical for cell proliferation and thus is naturally often connected to cancers. Cellular responses to stress pathway (R-HSA-2262752) is a subpathway of the cellular responses to external stimuli pathway (R-HSA-8953897) \cite{reactome}. Anticancer treatments are often successful when able to induce apoptosis through external stimuli that induce cellular stress \cite{herr2001cellular}. For instance, tumour suppressor gene P53 can be stimulated via cellular stress \cite{pflaum2014p53}. Thus, perturbation to those pathways might lead to cancer onset and resilience to treatment. 

The literature review highlights the ability of our classifier re-purposing approach to identify associations between biological pathways and cancer types that are supported by the existing literature. Thus, this analysis underlines the ability of our framework to extract cancer mechanisms at the level of biological pathways.

\section*{Discussion}

We introduce a two-step framework to perform data integration, feature reduction, and classification to uncover cancer-related knowledge. First, we develop an integrative non-negative matrix factorisation model to jointly embed entities in multiple connected latent spaces based on relational data between those entities. We show that relative positions of entities in our latent spaces are consistent with what we know about them. For instance, we show that genes group in functional domains and are close to associated higher-order molecular structures (pathways and complexes) embedded in the same latent space. Patients tend to be closer to other patients having the same diagnosis and to the diagnosis itself. By taking a pan-cancer approach, we are able to identify groups of patients with similar molecular manifestations spanning various cancers, confirming that cancer classification may need to be rethought on a global scale and the need for initiatives such as PCAWG \cite{the2020pan}. Based on known drug indications for the treatment of each cancer type and known cancer type driver genes, we train classifiers through which we can predict relevant new associations for each cancer type. Due to the joint embedding of different entities in the same latent space, we hypothesised that classifiers trained to identify associations with one type of entities could be re-purposed to derive associations with other, less-studied, entities. In this way, we can uncover biological mechanisms affected by each cancer type.

Interestingly, our work opens the door for actionable precision medicine. Through the joint embedding of cancers and patients, classifiers trained on high-level knowledge about cancer types can be re-purposed to help identify patient-specific information, such as potential drug treatment. However, as the biological validation of such predictions is difficult, requiring cell-line experiments or clinical trials, we leave it for future work.

Our framework is general and flexible and can accommodate additional and different data. While we focus on cancer here, our work paves the way for general cross-diseases analysis that could be useful to identify treatment re-purposing based on molecular similarities among medical conditions. 

\section*{Methods}

\subsection*{Data source and processing}\label{data_method}

We downloaded protein-protein interactions (PPI) data from Biogrid (version 3.5.176). We only keep interactions that have been validated experimentally using yeast-to-hybrid or affinity capture techniques. We obtained protein complexes data from CORUM and Reactome databases (both accessed in April 2019). Reactome is also used to collect all existing pathways of which we only keep pathways that have a traceable author statement (TAS). We further remove disease pathways which are only relevant in the associated disease context.

Patients data are obtained from the DCC Data release of the International Cancer Genome Consortium. We collected patients from $21$ cancer cohorts from TCGA studies (see Table \ref{tab:cancers}) and kept patients that have RNA-sequencing data, which adds up to 7,998 patients. We consider all Single Nucleotide Variations (SNV) reported in the data releases. 

\begin{table}[ht]
    \centering
    \begin{tabular}{p{4cm} c c || p{4cm} c c }
    \hline
        Cancer & Cohort size & Abbreviation & Cancer & Cohort Size & Abbreviation  \\
        \hline
         \rowcolor{black!10}Acute Myeloid Leukemia & 173 & LAML & Bladder Urothelial Carcinoma & 295 & BLCA  \\
         Brain Lower Grade Glioma & 439 & LGG & Breast invasive carcinoma & 1,041 & BRCA \\
         \rowcolor{black!10}Cervical squamous cell carcinoma and endocervical adenocarcinoma & 259 & CESC &  Colon adenocarcinoma & 428 & COAD \\
         Glioblastoma multiforme & 159 & GBM & Head and Neck squamous cell carcinoma & 480 & HNSC \\
         \rowcolor{black!10}Kidney renal clear cell carcinoma & 518 & KIRC & Kidney renal papillary cell carcinoma & 222 & KIRP \\
         Liver hepatocellular carcinoma & 173 & LIHC & Lung adenocarcinoma & 477 & LUAD \\
         \rowcolor{black!10}Lung squamous cell carcinoma & 428 & LUSC & Ovarian serous cystadenocarcinoma & 262 & OV \\
         Pancreatic adenocarcinoma & 142 & PAAD & Prostate adenocarcinoma & 375 & PRAD \\
         \rowcolor{black!10}Rectum adenocarcinoma & 153 & READ & Skin Cutaneous Melanoma & 430 & SKCM \\
         Stomach adenocarcinoma & 415 & STAD & Thyroid carcinoma & 500 & THCA \\
         \rowcolor{black!10}Uterine Corpus Endometrial Carcinoma & 508 & UCEC & & & \\
         \hline
         
    \end{tabular}
    \caption{List of cancer types considered in this study with associated abbreviations from TCGA.}
    \label{tab:cancers}
\end{table}{}
 
We consider the set of 15,224 genes whose transcripts are measured for by the RNA-sequencing technology across all datasets and that have at least one PPI with another selected gene according to BioGrid data. We derive a gene expression vector from RNA-seq measurement for each patient that is normalised to Transcripts Per Million (TPM) and rescaled using logarithm in base $2$, i.e. the expression score of a gene is given by $\log_2(\textnormal{TPM}+1)$. Note that $559$ patients have do not have any mutation on any of the 15,224 genes considered.

Drug--target and chemical data is obtained from DrugBank (version 5.1.3) \cite{wishart2017drugbank}. We consider all drugs that are approved, experimental, or investigational. Drugs chemical similarity is computed using the Tanimoto similarity \cite{nikolova2003approaches} between drugs smile representations. The details of the data used can be found in Table \ref{tab:data}.
\begin{table}[ht]
    \centering
    \begin{tabular}{c c c c}
    \hline
        Data & Size  & Density & Symbol \\
        \hline
         Gene expression &  7,998$\times$15,224 & n.a & $X_{exp}$\\
         Gene SNV &  7,998$\times$15,224 & $1.7\%$ & $X_{snv}$\\
         Patient cancer type &  21$\times$7,998 & n.a. & $X_{pct}$\\
         PPI &  15,224$\times$15,224 & $0.21\%$ & $X_{ppi}$\\
         Protein complexes &  7,022$\times$15,224 & $0.05\%$ & $X_{pc}$\\
         Biological pathways &  1,650$\times$15,224 & $0.32\%$ & $X_{bp}$\\
         Drug--target &  7,333$\times$15,224 & $0.013\%$ & $X_{dt}$\\
         Drug tanimoto similarity &  7,333$\times$7,333 & n.a. & $X_{ts}$\\
         \hline
    \end{tabular}
    \caption{Details of the data used. The columns correspond to: 1) the type of data, 2) the size of the matrix representing the data, 3) the density, where applicable, indicates the percentage of the existing links between the entities out of all possible links, and 4) the symbol used in the document to refer to the matrix containing the corresponding data.}
    \label{tab:data}
\end{table}{}

Genes' annotations used for the enrichment analysis are obtained from Gene Ontology (GO)\cite{ashburner2000gene} (release 10/06/2019). We keep annotations that have an experimental evidence code (one of EXP, IDA, IPI, IMP, IGI, and IEP). We consider all three GO annotation subtypes separately: Biological Processes (GO--BP), Molecular Function (GO--MF) and Cellular Component (GO--CC). For each, we build a directed acyclic graph (DAG) that connects annotations based on  ``is a'' relationships (we use the go-basic.obo file giving annotations relationships available on GO's website). Then, we propagate the annotations for each gene up the corresponding DAG, which means that we add to the set of annotations of a gene the union of ancestors of the annotations. We remove annotations that annotate less than $0.1\%$, or more than $10\%$ of the 15,224 genes considered, i.e. we prune annotations that are either too rare, or too common. We give the statistics of annotations in Table \ref{tab:go}

\begin{table}[ht]
    \centering
    \begin{tabular}{c || c c c}
    \hline
        GO subtype & GO--BP  & GO--MF & GO--CC \\
        \hline
        Number of annotations & 2,322 & 538 & 366 \\
        Percentage of genes annotated & $50\%$ & $43\%$ & $45\%$\\
         \hline
    \end{tabular}
    \caption{GO annotations statistics for all GO subtypes.}
    \label{tab:go}
\end{table}{}

\subsection*{Non-negative Matrix Factorizations}\label{nmf_method}

Matrix factorizations approaches aim to approximate a matrix $X$ by the product of $n$ smaller matrices $F_i,\-i\in\{1..n\}$, called factors, i.e. $X \approx \prod_i F_i$. Mathematically, this amounts to finding factors $F_i$, under user defined dimensional contraints, that minimize the equation $\|X -  \prod_i F_i\|^2_F$, where $\|\cdot\|_F$ represents the Frobenius norm of a matrix. Non-negative matrix factorizations techniques add a positivity contraint on the factors, i.e. $\forall i, \-F_i \geq 0$.

The objective is to obtain lower dimensional representation that captures the essence of the data and can be used to identify missing entries through the matrix completion property. In this work, we use three variants of non-negative matrix factorisations approaches. \\

\noindent\textbf{NMF} decomposes a rectangular matrix $X\in \mathbb{R}^{m\times n}$ in the product of two positive factors $F\in\mathbb{R}_+^{m\times k}$ and $G\in\mathbb{R}_+^{n\times k}$, with $k\leq\min(m,n)$, such that $\|X - FG^T\|^2_F$ is minimized. With NMF, the embeddings, given by $F$ and $G$, of the two groups of entities whose relational data is given by $X$, are in the same latent space.\\

\noindent\textbf{NMTF} decomposes a rectangular matrix $X\in \mathbb{R}^{m\times n}$ in the product of three positive factors $F\in \mathbb{R}_+^{m\times k_1}$, $S\in \mathbb{R}_+^{k_1\times k_2}$ and $G\in \mathbb{R}_+^{n\times k_2}$, with $k_1,k_2\leq\min(m,n)$, such that $\|X - FSG^T\|^2_F$ is minimized. \\

\noindent\textbf{SNMTF} decomposes a symmetric matrix $X\in \mathbb{R}^{n\times n}$ in the product of two positive factors $G\in \mathbb{R}_+^{n\times k}$ and $S\in \mathbb{R}_+^{k\times k}$, with $k\leq n$, such that $\|X - GSG^T\|^2_F$ is minimized. 

\subsection*{Our Integrative Framework}\label{integrative_framework}

We propose a framework that jointly integrates all data sources together by using a mixture of non-negative matrix factorisations to obtain for each entity an embedding that takes into account the full context of our task. We minimize the following general objective function $\mathcal{L}$ over all $G_\cdot$ and $S_\cdot$ factors:

    \begin{align} \label{objective}
    \mathcal{L} = \min_{G_\cdot,S_\cdot} &\sum_{x\in\{exp,snv\}}\|X_{x} - G_{p}S_{x}G_{g}^T\|^2_F + \sum_{x \in \{pc,bp\}}\|X_x - G_{x} G_{g}^T\|^2_F \\
    &+  \|X_{ppi} - G_{g}S_{ppi}G_{g}^T\|^2_F + \|X_{pct} - G_{ct}G_{p}^T\|^2_F\nonumber\\ 
    &+\|X_{ts} - G_{d}S_{ts}G_{d}^T\|^2_F +\|X_{dt} - G_{d}G_{g}^T\|^2_F\nonumber,
    \end{align}  

\noindent where, henceforth, each $X_{D}$ represents the matrix associated to data type $D$, see nomenclature in Table \ref{tab:data}, each $G_{E}$ factor gives the embeddings of the entities of type $E$, with subscripts $g$, $p$, $ct$, $d$, $bp$, and $pc$ corresponding, respectively, to genes, patients, cancer types, drugs, pathways, and complexes. $S_\cdot$ factors are optimized over, but not used for the analysis.

The integration of the various data sources is achieved by sharing factors across the NMF sub-objectives that constitute our global objective function $\mathcal{L}$. For instance, the factor $G_g$, corresponding to the genes embeddings, is shared by all decompositions that involve genes which corresponds to the factorisation of PPI data ($X_{ppi}$), the factorisations of patients molecular data ($X_{exp}$ and $X_{snv}$), the factorisation of drug--target data ($X_{dt}$), and the factorisations of higher-order biological entities ($X_{pc}$ and $X_{bp}$). Through this factor sharing and joint optimisation, the framework is able to harness the relevant information contained across the data sources to derive meaningful embeddings. 

\subsection*{Optimization}\label{optimization}
The minimisation of the objective function given in Equation \ref{objective} is achieved through an iterative optimisation process. We use in our framework multiplicative update rules \cite{lee2001algorithms} designed to maintain non-negativity of all the factors in the decomposition.

We use an initialization strategy based on the truncated singular value decomposition (SVD) for all factors that has shown better performances than random initialization \cite{boutsidis2008svd,malod2019towards} and has the advantage of giving deterministic solutions. Specifically, consider a factor $G\in \mathbb{R}^{n\times k}$ involved in the decomposition of $l$ data matrices $X_i,\ i\in\{1..l\}$. Without loss of generality, we assume that $G$ is the right hand side factor in the decompositions, i.e. $\forall i,\-X_i \approx G F_i$. We denote by $U_i$ the right hand side term in the SVD decomposition of $X_i$ ( $X_i = U_i S_i V_i^T$ ). We introduce $U_i^+ = \max(U_i,0)$ and $U_i^- = \max(-U_i,0)$. We then denote by $\tilde{U}_i\in \mathbb{R}^{n\times k}$ the matrix where each column is defined by
\begin{align}
\tilde{U}_i(j) = 
\begin{cases}
    \sqrt{s_i(j)}U_i^+(j),& \text{if } \|U_i^+(j)\|\geq \|U_i^-(j)\|\\
    \sqrt{s_i(j)}U_i^-(j),              & \text{otherwise}
\end{cases},\nonumber
\end{align}{}

\noindent where $M(j)$ denotes the $j^{th}$ column of the matrix $M$ and $s_i(j)$ is the $j^{th}$ largest singular value of $X_i$. Factor $G$ is then initalized as $\frac{1}{l}\sum_{i=1}^{l}\tilde{U}_i + \epsilon$. We initialize  the central matrix in NMTF decompositions to $I+\epsilon$, where $I$ denotes the identity matrix. Under the multiplicative update rules, any entries initialized to zero would stay null. Hence, we add a small $\epsilon$ everywhere to allow all entries to vary.

The iterative optimisation is ran either for $200$ epochs or until the relative variation of the objective function between two consecutive epochs is lower than $10^{-4}$, i.e. when $\frac{|\mathcal{L}_{t+1}-\mathcal{L}_t|}{\mathcal{L}_t} \leq 10^{-4}$ where $\mathcal{L}_t$ corresponds to the value of the objective function at iteration $t$.
 
\subsection*{Boosted decision tree}
In the main document, we used boosted decision trees to predict cancer type associations with entities that are part of our framework based on the embeddings derived from the Joint NMF optimisation step. A decision tree classifier partitions the input data iteratively based on features. Boosting signifies deriving a strong classifier from the serial associations of weak classifiers. In our case, the base classifiers are decision trees. The boosted algorithm iteratively adds decision trees to the classifier with the aim of reducing the error of the previous classifier\cite{roe2006boosted}. 

 We discuss here the implementation details that we used. First, we use boosted decision trees from the xgboost package \cite{chen2016xgboost}. Boosted decision trees have different hyperparameters that control various aspect of the algorithm: $\eta$ controls the learning rate, $\gamma$ corresponds to a threshold under which a leaf node of the decision tree is not split anymore, the maximal depth of a decision tree, and $\lambda$ controls the L2-regularization (for more details see \cite{chen2016xgboost}). We perform a $10$-fold cross-validation to fix those hyperparameters with $\eta\in\{0.25,0.5,0.75\}$, $\gamma\in\{0,10\}$, max depth$\in\{6,12\}$, and $\lambda\in\{1,10,100\}$. The best set of parameters is chosen as the one that leads to the classifier with the highest AUROC score in the associated task. Note that we also use early-stopping during training with an $80\%$/$10\%$/$10\%$ train/validation/test split of our data. We use all $10$ classifiers trained during the cross-validation process to derive an association score for each possible pair. To ensure that the scoring of the $10$ classifiers is comparable, we rescale the output scores to have $0$ mean and unit variance. The average of all classifier scores then gives the final association score of an entity pair. 

\subsection*{Baselines}

We contrast the performances of our trained boosted decision tree with those of the state-of-the-art methods for the prediction of cancer type associations with genes and drugs. We chose baselines based on the availability of source code (or detailed implementation description), the quality of reported performances, and the concordance of input data with ours. Our implementation of each method is available in the Supplementary Files. When a method requires hyperparameters tuning, the criterion used to identify the best set of hyperparameters is always the AUROC score of the classifier in the associated task.

\paragraph{Non-negative Matrix Factorization Reconstruction (NMFR)}  is based on the reconstruction of the data after factorizations and is the simplest approach based on our framework. The idea is based on the matrix completion property observed in matrix factorizations methods\cite{mehta2017review}. Here, we propose a simple method that makes use of the link between factors to extract entities' association scores. For instance, cancer--gene association scores, $CG$, are given by \[ CG = G_{ct} G_p^T G_p \frac{\left( S_{exp}+S_{snv} \right)}{2}G_{g}^T,\] where entry $(i,j)$ of matrix $CG$ gives the association score between cancer type $i$ and gene $j$. Cancer--drug association scores, $CD$, are given by \[ CD = G_{ct} G_p^T G_p \frac{\left( S_{exp}+S_{snv} \right)}{2}G_{g}^TG_{g}S_{dt}G_{d}^T,\]where entry $(i,j)$ of matrix $CD$ gives the association score between cancer type $i$ and drug $j$.

Performances are measured by how well those association scores correlate to IntOGen cancer--driver data and DrugCentral cancer--drug data using AUROC and AUPRC.

\paragraph{MBiRW}\cite{luo2016drug} was proposed to identify potential new indications for the existing drugs. The method is based on a bi-directional random walk using a drug similarity network, a disease similarity network, and a bipartite network connecting diseases to drugs. The authors report good performances against known ground-truth relative to the competing methods and manually validate de novo predictions.

Here, the drug similarity network adjacency matrix is given by the drug Tanimoto similarities matrix $X_{ts}$. We define the cancer similarity network $X_{cc}$ based on a molecular similarity between cancers. We first associate to each cancer two molecular signatures given by the average of patients gene expression data and SNV data. From each type of data, we define a cancer similarity network that corresponds to the cosine similarity of their molecular signatures. We denote by $X_{cc}$ the final cancer similarity network corresponding to the average of those two similarity networks. Note that the authors use a different disease similarity matrix, the source of which is currently offline. Finally, we use cancer--drug data from DrugCentral to define a bipartite network in which an entry is set to $1$ to indicate an association between the corresponding drug and cancer, and $0$ otherwise. 

The authors propose an iterative method that follows the step given in Algorithm \ref{alg:mbirw}.

\begin{algorithm}[ht]
 \KwData{cancer--cancer network $X_{cc}$, drug--drug network $X_{ts}$, cancer--drug network $CD$, parameter $\alpha$, maximum number of iterations $M$ }
 \KwResult{cancer--drug associations scores $O$}
 $DD = D_{ts}^{-\frac{1}{2}}X_{ts}D_{ts}^{-\frac{1}{2}}$;where $D_{ts}$ is a diagonal matrix where entry $D_{ts}(i,i) = \sum_j X_{ts}(i,j)$\\
 $CC = D_{cc}^{-\frac{1}{2}}X_{cc}D_{cc}^{-\frac{1}{2}}$;where $D_{cc}$ is a diagonal matrix where entry $D_{cc}(i,i) = \sum_j X_{cc}(i,j)$\\
 $R_0 = \frac{CD}{\sum CD}$; where $\sum CD$ gives the number of non-zero entries in $CD$\\
 $O = R_0$; $n_iter = 0$;\\
 \While{$n_{iter} \leq M$}{
  $L = \alpha\cdot  O\cdot DD + (1-\alpha)\cdot  R_0$;\\
  $R = \alpha \cdot CC\cdot O + (1-\alpha)\cdot  R_0$;\\
  $O = \frac{L + R}{2}$;\\
  $n_{iter}=n_{iter}+1;$
  }
 \caption{MBiRW algorithm.}
 \label{alg:mbirw}
\end{algorithm}

We perform a $10$-fold cross-validation to select hyperparameters $\alpha \in \{0.1,0.3,0.5,0.7,0.9\}$ and $M \in \{2,5,10,20\}$ (note that the authors set $M=2$, and search for $\alpha$). In each run, $10\%$ of known cancer--drug associations are masked in the input to the algorithm and we evaluate how well MBiRW is able to retrieve those. 

\paragraph{DRRS}\cite{luo2018computational} was proposed to identify potential new indications for the existing drugs as well. The method is based on the matrix completion property of Singular Value Thresholding Algorithm (SVT) using a drug similarity matrix, a disease similarity matrix, and a disease--drug indication matrix. The authors report good performances against known ground-truth relative to the competing methods and further use their methods to predict indications for new drugs, validating novel associations. 

For our purposes, we use the Tanimoto drug similarity matrix $X_{ts}$, the cancer--cancer similarity matrix derived from ICGC molecular data $X_{cc}$, and cancer--drug associations from DrugCentral $X_{cd}$. The authors define the block matrix $A$

\[A = \left(
\begin{array}{c c}
X_{ts} & X_{cd}^T \\

X_{cd} & X_{cc}
\end{array}
\right),\]
and feed it to the SVT algorithm. The maximum number of epochs is set to the minimum between the number of cancers and the number of drugs. The iteration with the highest AUROC gives the final predictions. We perform a $10$-fold cross-validation to evaluate the performance of the algorithm. In each iteration, we mask $10\%$ of known cancer--drug associations and evaluate how well the algorithm retrieves them. 

\paragraph{BNNR}\cite{yang2019drug} was developed for re-purposing of drugs. The method is also based on the completion property of SVT. The algorithm follows the steps given in Algorithm \ref{alg:bnnr}. Compared to DRRS, BNNR incorporates a regularisation term to balance the approximation error and the rank properties and thus can handle the noisy drug–drug and disease–disease similarities. It also adds a constraint that clips the association scores to the interval $[0,1]$. The authors found that those additions benefited performances compared to DRRS for retrieval of known associations. They further manually validated the top-scoring associations through literature curation. 

\begin{algorithm}[ht]
 \KwData{block matrix $A$, hyperparameters $\alpha$ and $\beta$}
 \KwResult{cancer--drug associations scores $W$}
 $tol_1 = 2\cdot 1e^{-3}$; $tol_2 = 1e^{-5}$; \\
 $M = 300$; $n_{iter} = 0$;\\
 $s_1 = 1$; $s_2 = 1$; \\
 $X = A$; $Y = X$; $W = X$;\\
 $mask = T > 0$;\\
 \While{$n_{iter} \leq M$ or $tol_1 < s_1$ or $tol_2 < s_2$ }{
    $F = \frac{1}{b}(Y + \alpha T) + X$;\\
    $W = F - \frac{\alpha}{\alpha+\beta} F \odot mask $;\\
    $W[W < 0] = 0$; $W[W>1]=1$; \\
    $X_t = \textnormal{svt}(W-\frac{1}{\beta}Y,\frac{1}{\beta})$;\\
    $Y = Y + \beta (X_t - W)$;\\
    $s_t = s_1$; $s_1 = \frac{\|X_t-X\|_F}{\|X\|_F}$;\\
    $s_2 = \frac{|s_1-s_t|}{\max(1,|s_t|)}$;\\
    $X = X_t$; $n_{iter} = n_{iter} +1$;
  } 
 \caption{BNNR algorithm.}
 \label{alg:bnnr}
\end{algorithm}
We perform a $10$-fold cross-validation to evaluate the performance of the algorithm and fix the hyperparameters $\alpha\in\{0.1,1,10,100\}$ and $\beta\in\{0.1,1,10,100\}$. 

\paragraph{Network Based Integration (NBI)}\cite{ruffalo2015network} was developed to identify cancer-related genes that are not necessarily mutated or differentially expressed. The method is based on network heat diffusion process over a molecular network. The original paper focuses on a specific cancer for which they collect differential gene expression data and SNV data. The authors assess performances by first measuring how accurately their method retrieves known cancer driver genes and then validate novel cancer--gene associations.

Network heat diffusion is defined by the iterative update of scores $X^0$ associated to the network's nodes following the equation  \[ X^{n+1} = \alpha W X^{n} +(1-\alpha)X^0,\] where $W$ denote the network data and $X^n$ corresponds to the updated scores after $n$ iterations. The iterative process terminates when $\|X^{n+1}-X^n\|_2 < 10^{-6}$.

The authors set $W$ to the PPI normalized adjacency $W= \Delta^{-\frac{1}{2}} X_{ppi} \Delta^{-\frac{1}{2}}$, where $\Delta^{-\frac{1}{2}}$ is a diagonal matrix where entry $i$ corresponds to the degree of gene $i$ in the PPI network. The authors use the diffusion process both on patients differential gene expression and SNVs, obtaining two diffused vector scores per patient. They then handcraft $13$ cancer-specific features for each gene based on the results. Those features are then used as input to a logistic regression classifier trained to predict known cancer drivers.

As differential gene expression is not available to us, we use the same gene expression values that are input to our framework. Since our analysis is across cancers, we compute gene features for each cancer types, i.e. each cancer--gene pair is associated with a $13$-dimensional feature vector. The method has two hyperparameters: $\alpha$ for the heat diffusion process, and $C$ that controls regularisation of the logistic classifier. We perform a $10$-fold cross-validation procedure to pick the best pair of hyperparameters, with $\alpha\in\{0.25,0.5,0.75\}$ and $C\in\{0.01,1,100\}$, and to evaluate the performance of a logistic regression classifier trained on those features to predict cancer-specific driver genes.

\paragraph{LOTUS}\cite{collier2019lotus} is a method that achieved the state-of-the-art results for the more specific tasks of identifying oncogenes and tumour-suppressing genes. Each task is tackled separately, and each with $3$ different gene features that are not available to us. However, the method can be adapted to the simpler task of retrieving cancer driver genes. To this end, we use gene expression data, SNVs, and gene methylation data, that are available in ICGC, as gene features. A patient's gene methylation is defined as the average beta value of all associated CpG islands. 

The authors of LOTUS propose both a cancer-specific framework and a pan-cancer framework; we use the latter here. The method revolves around the Support Vector Machine (SVM) algorithm. The authors first define for each sample--gene pair $3$ features that are then averaged across all samples to give the final gene features. In their work, the final features correspond to the number of damaging missense mutations, the total number of missense mutations, and the entropy of the spatial distribution of the missense mutations on each gene, for the prediction of oncogenes. For the prediction of tumour-suppressing genes, the features are the number of frameshift mutations, the number of loss-of-function mutations, and the number of splice site mutations. Note that when defining those features, the authors do not differentiate across cancer types. In our case, the features correspond to the mutation frequency, the average gene expression, and the average gene methylation across all samples. To ensure that those features are comparable, we normalise the distributions to have $0$ mean and unit variance. 

The authors then define both a gene kernel $K_{g}$ and a cancer kernel $K_{c}$. The gene kernel is defined as the average of a kernel corresponding to a gene similarity matrix derived from $3$-dimensional features defined above and a kernel derived from the PPI network. We have \[K_g = \frac{1}{2}\left(\Phi\Phi^T + e^{-L}  \right),\] where $\Phi\in\mathbb{R}^{15,224\times 3}$ represents the gene features and $L$ is the normalized Laplacian of the PPI network, $L = I - D^{-\frac{1}{2}}X_{ppi}D^{-\frac{1}{2}}$, $I$ represents the identity matrix and $D$ the diagonal matrix with entries corresponding to the degree of each node in the PPI network. The cancer kernel is defined as the sum of three kernels\[K_c = \frac{1}{3}\left(I + J + X_{cc} \right),\] where $I$ represents the identity matrix, $J$ corresponds to the matrix filled with ones, and $X_{cc}$ is a cancer similarity matrix. As above, we use the cancer similarity matrix defined based on cancers molecular similarities.

The final kernel $K$ for (cancer,gene) pairs used for pan-cancer analysis is defined by \[K\left((c,g),(c',g')\right) = K_{c}(c,c') \times K_{g}(g,g'),\] where $c$ and $c'$ represent cancers and $g$ and $g'$ represent genes. The hyperparameter of the model corresponds to the regularisation coefficient $C$ of the SVM algorithm. Due to the large size of the full kernel $K$, we use the same strategy as the authors of LOTUS and randomly sample negative (cancer,gene) pairs from all (cancer,gene) pairs that are not reported in IntOGen. This effectively boils down to using a submatrix of kernel $K$ as input that contains as many positive (cancer--driver associations) and negative pairs. As before, we perform cross-validation to pick $C\in\{1,10,100,1000\}$ and evaluate the performance of the method on our task.

\paragraph{Subdyquency}\cite{song2019random} is a method based on random walks on a network to identify cancer drivers. It achieves the state-of-the-art results for the retrieval of known cancer drivers. The framework is defined for specific cancers and we extend it to pan-cancer. In their framework, the authors build a network between ``outlier'' and mutated genes. The outlier genes are genes whose expression is significantly different with respect to the cohort. They correspond to genes with absolute z-score strictly greater than $2$. The set of outlier (mutated) genes is defined as all genes being at least outlier (mutated) for one patient. Here, we consider all cancers together to define those sets. Directed interactions between genes are obtained from the Functional Interactions (FI) network \cite{wu2010human} derived from Reactome \cite{reactome}. We downloaded the 2019 version of the FI network. The authors define a bipartite graph between the two sets of genes whose edges correspond to directed links in the FI network. The edge weights are defined based on the localisation of proteins in a cell as given by the COMPARTMENT database \cite{binder2014compartments} (see the original paper for details). As done by the authors, we downloaded all data relating to human regardless of evidence type (obtained in April 2020). We denote the adjacency matrix of this bipartite graph with $W\in\mathbb{R}^{n_m\times n_o}$, where $n_m$ and $n_o$ represent the number of mutated and outlier genes, respectively. Then, for each patient $p$, the authors define a feature vector for the outlier gene set, denoted by $O_p$, and another one for the mutated gene set, denoted by $M_p$. Specifically, consider gene $i$ in the mutated set. If gene $i$ is mutated for patient $p$, then $M_p(i)$ is set to the mutation frequency of gene $i$ in the cohort of patients having the same cancer as $p$, and $0$ otherwise. For gene $j$ in the outlier set, if $j$ is not an outlier for patient $p$, then $O_p(j)$ is set to $0$, if $j$ is an outlier and is also in the mutated set, then $O_p(j)$ is set to $M_p(j)$, else it is set to the outlier frequency of $j$ across the set of patients having the same cancer as patient $p$. The authors then propose the three steps procedure simulating a random walk on the bipartite graph using both feature vectors
\begin{align*}
    R_p^m &= \alpha M_p + (1-\alpha) W O_p,\\
    R_p^o &= \alpha O_p + (1-\alpha) W^T R_p^m,\\
    R_p^m &= \alpha M_p + (1-\alpha) W R_p^o,
\end{align*}{}
\noindent where $\alpha\in[0,1]$ is the sole hyperparameter of the model. The final cancer--gene scores are derived by summing the $R_p^m$ vectors across patients. Higher scores indicate a stronger associations between a cancer and a gene. We perform a cross-validation to pick $\alpha\in [0,1]$ and evaluate the performance of the method on our task.

\section*{Data Availability}
The data is available at \url{https://life.bsc.es/iconbi/context_aware_embeddings/index.html}.

\section*{Code Availability}

The c++ library for the joint factorisation is available at \url{github.com/tgaudele/nmfif}.

\noindent The analysis code, including baselines methods, is available at \url{https://life.bsc.es/iconbi/context_aware_embeddings/index.html}.

\bibliography{sample}

\section*{Funding}
This work was supported by the European Research Council (ERC) Consolidator Grant 770827 and UCL Computer Science.

\section*{Author contributions statement}
T.G. contributed to developing the computational concepts, implementation, data analysis, results interpretation, and manuscript writing. N.M.D. contributed to developing the computational concepts, results interpretation, and manuscript writing. N.P conceived, directed and supervised the project, contributed to results interpretation and to manuscript writing. 

\section*{Additional information}
The authors declare no competing interests.


\end{document}